\documentclass[10pt,conference]{IEEEtran}

\usepackage{amsfonts}
\usepackage[dvips]{graphicx}
\usepackage{times}
\usepackage{cite}
\usepackage{amsmath}
\usepackage{cases}
\usepackage{array}
\usepackage{dsfont}
\usepackage{amssymb}

\usepackage{stfloats}
\usepackage{slashbox}
\usepackage{graphicx}
\usepackage{footnote}
\usepackage{booktabs}
\usepackage{array}
\usepackage{bm}

\begin{document}

\title{A New DoF Upper Bound and Its Achievability for $K$-User MIMO Y Channels}
\author{\IEEEauthorblockN{Kangqi Liu and Meixia Tao}
\IEEEauthorblockA{Dept. of Electronic Engineering, Shanghai Jiao Tong University,  Shanghai, China\\
Emails: \{forever229272129, mxtao\}@sjtu.edu.cn}
}

\maketitle

\begin{abstract}
This work is to study the degrees of freedom (DoF) for the $K$-user MIMO Y channel. Previously, two transmission frameworks have been proposed for the DoF analysis when $N \geq 2M$, where $M$ and $N$ denote the number of antennas at each source node and the relay node respectively. The first method is named as signal group based alignment proposed by Hua \textit{et al}. in \cite{Mu1}. The second is named as signal pattern approach introduced by Wang \textit{et al}. in \cite{Wang3}. But both of them only studied certain antenna configurations. The maximum achievable DoF in the general case still remains unknown. In this work, we first derive a new upper bound of the DoF using the genie-aided approach. Then, we propose a more general transmission framework, generalized signal alignment (GSA), and show that the previous two methods are both special cases of GSA. With GSA, we prove that the new DoF upper bound is achievable when $\frac{N}{M} \in \left(0,2+\frac{4}{K(K-1)}\right] \cup \left[K-2, +\infty\right)$. The DoF analysis in this paper provides a major step forward towards the fundamental capacity limit of the $K$-user MIMO Y channel. It also offers a new approach of integrating interference alignment with physical layer network coding.
\end{abstract}


\section{Introduction}
Degrees of freedom (DoF) characterize the independent number of data streams the system can transmit and is thus a measure of the asymptotic capacity at high signal-to-noise ratio (SNR)\cite{Cadambe}. Interference alignment (IA) is an attractive technique to manage the interference and enhance DoF of various interference channels \cite{Jafar,Maddah}. The basic idea of IA is to align the interference signals in a same signal space so as to leave more signal space to transmit useful signals.

By integrating the concepts of IA and physical-layer network coding (PLNC), signal alignment (SA) was firstly proposed in \cite{Lee1} to analyze the achievable DoF of the MIMO Y channel, where three source nodes exchange independent messages with each other via a common relay node. By comparing with the traditional time-division multiple access (TDMA) scheme and multiuser-MIMO (MU-MIMO) scheme, the achievable DoF with SA is tripled and doubled, respectively. Later, the authors in \cite{Chaaban1} showed that the DoF upper bound of the MIMO Y channel is $\min\{3M,2N\}$, where $M$ and $N$ denote the number of antennas at each source node and the relay node respectively. The DoF upper bound $3M$ is proved to be achievable when $\frac{N}{M} \geq \frac{3}{2}$ in \cite{Lee1} and the DoF upper bound $2N$ is proved to be achievable when $\frac{N}{M} \leq \frac{3}{2}$ in \cite{Chaaban1}. Hence, the DoF analysis of the MIMO Y channel is completed with SA.

A natural generalization of the MIMO Y channel is $K$-user MIMO Y channel, where $K$ users exchange independent messages with each other via the relay. The DoF analysis of the $K$-user MIMO Y channel has attracted a lot of attention \cite{Lee,Wang3,Mu1,Wang4,Liu3,Liu5}.

\begin{table*}[tbp]
\tiny
\centering
\caption{Recent Advances towards the DoF Analysis}\label{table1}
\begin{tabular}{|c|c|c|c|}
\hline
$K$ & $\frac{N}{M}$ & Maximum DoF & Reference\\ \hline
3 &  $\left[\frac{3}{2}, +\infty\right)$ & $3M$ & \cite{Lee1}\\ \hline
3 &  $\left(0,+\infty\right)$ & $\min\{3M,2N\}$ & \cite{Chaaban1} \\\hline
4 &  $\left(0,\frac{12}{7}\right] \cup \left[\frac{8}{3}, +\infty\right)$ & $\min\{4M,2N\}$ & \cite{Yuan1} \\\hline
4 &  $\left(0,\frac{12}{7}\right] \cup \left[\frac{7}{3}, +\infty\right)$ & $\min\{4M,2N\}$ & \cite{Liu5} \\\hline
4 &  $\left(0,+\infty\right)$ & $\max\{\min\{4M,\frac{12N}{7}\},\min\{\frac{24M}{7},2N\}\}$ & \cite{Wang4} \\\hline
$K>4$ &  $\left(0,\frac{2K^2-2K}{K^2-K+2}\right]$ & $\min\{4M,2N\}$ & \cite{Lee} \\\hline
$K>4$ &  $\left(K-1,+\infty\right]$ & $KM$ & \cite{Mu1} \\\hline
$K>4$ &  $\left(0,\frac{2K^2-2K}{K^2-K+2}\right] \cup \left[\frac{K^2-2K}{K-1}, +\infty\right)$ & $\min\{KM,2N\}$ & \cite{Wang3} \\ \hline
$K>4$ &  $\left(0,\frac{2K^2-2K}{K^2-K+2}\right] \cup \left[\frac{K^2-3K+3}{K-1}, +\infty\right)$ & $\min\{KM,2N\}$ & \cite{Liu5} \\ \hline
$K>4$ &  $\left(0,2+\frac{4}{K(K-1)}\right] \cup \left[K-2, +\infty\right)$ & \eqref{dof_KY_upper_information} & This paper \\ \hline
\end{tabular}
\end{table*}

\textsc{Table} I summarizes the recent advances towards the DoF analysis of the $K$-user MIMO Y channel. In particular, the DoF analysis when $K=3$ and $K=4$ is completed with \cite{Chaaban1} and \cite{Wang4}. The maximum achievable DoF when $K>4$ with the antenna configuration $\frac{N}{M} \in \big(\frac{2K^2-2K}{K^2-K+2}, \frac{K^2-3K+3}{K-1}\big)$ is still an open problem.

In this paper, we are interested in the DoF analysis of the $K$-user MIMO Y channel for the antenna configuration $\frac{N}{M} \in \big(\frac{2K^2-2K}{K^2-K+2}, \frac{K^2-3K+3}{K-1}\big)$. We first derive a new DoF upper bound and then we analyze its achievability. Finally, we show that the maximum DoF of the $K$-user MIMO Y channel is achieved under the antenna configuration $\frac{N}{M} \in \big(0, 2+\frac{4}{K(K-1)}\big] \cup \big[K-2, +\infty\big)$.

The main results obtained in this work are as follows:
\begin{itemize}
  \item A new DoF upper bound is derived for the $K$-user MIMO Y channel.
  \item We extend the generalized signal alignment method in \cite{Liu5} to the case when $\beta \neq K-2$ so that the DoF achievability is closer to the DoF upper bound.
  \item It is proved that the DoF capacity of the $K$-user MIMO Y channel is achieved under the antenna configuration $\frac{N}{M} \in \big(0, 2+\frac{4}{K(K-1)}\big] \cup \big[K-2, +\infty\big)$.
\end{itemize}

Notations: $(\cdot)^{T}$ and $(\cdot)^{H}$ denote the transpose and the Hermitian transpose, respectively. tr({\bf X}) and rank({\bf X}) stand for the trace and rank of {\bf X}. $\varepsilon[\cdot]$ stands for expectation. $\mbox{span} ({\bf X})$ and ${\mbox{null} ({\bf X})}$ stand for the column space and the null space of the matrix ${\bf X}$, respectively. $\mbox{dim}({\bf X})$ denotes the dimension of the column space of ${\bf X}$. $\lfloor x \rfloor$ denotes the largest integer no greater than $x$. $\lceil x \rceil$ denotes the smallest integer no less than $x$. {\bf I} is the identity matrix. $\left[{\bf X}\right]_{i,j}$ denotes the $(i,j)$-th entry of the matrix $\bf X$.

\section{Channel Model}
We consider the $K$-user MIMO Y channel. It consists of $K$ source nodes, each equipped with $M$ antennas, and one relay node, equipped with $N$ antennas. Each source node exchanges independent messages with all the other source nodes via the relay. It is assumed that the users can communicate only through the relay and no direct links exist between any pairs of users. All the source nodes in the network are assumed to be full duplex. The independent message transmitted from source node $i$ to source node $j$ is denoted as $W_{i,j}$. At each time slot, the message $W_{i,j}$ is encoded into a $d_{i,j} \times 1$ symbol vector $\textbf{s}_{i,j}=[s_{i,j}^1,s_{i,j}^2,\cdots,s_{i,j}^{d_{i,j}}]^T$, where $d_{i,j}$ denotes the number of independent data streams transmitted from source $i$ to source $j$.

The communication of the total messages takes place in two phases: the multiple access (MAC) phase and the broadcast (BC) phase. In the MAC phase, all $K$ source nodes transmit their signals to the relay simultaneously. Let ${\bf x}_i$ denote the transmitted signal vector from source node $i$. It is given by
\begin{eqnarray}\label{x_1}
{{\bf x}}_i =\sum\limits_{j =1,j \neq i}^{K} {\bf V}_{i,j}{\bf s}_{i,j}= {\bf V}_i{\textbf{s}}_i,
\end{eqnarray}
where ${\bf V}_{i,j}$ is the $M \times d_{i,j}$ precoding matrix for the information symbol vector ${\bf s}_{i,j}$ to be sent to source node $j$, ${\bf V}_i$ is a matrix obtained by stacking $\{{\bf V}_{i,j} \mid j \neq i\}$ by column and ${\bf s}_i$ is a vector obtained by stacking $\{{\bf s}_{i,j} \mid j \neq i\}$ by rows. Each transmitted signal ${\bf x}_i$, for $i=1,~\cdots,~K$, satisfies the power constraint of
\begin{equation}\label{Power_constraint_source}
\textrm{tr}({\bf x}_i{\bf x}_i^H) \leq P_{\rm{s}},
\end{equation}
where $P_{\rm{s}}$ is the maximum transmission power allowed at each source node.

The received signal ${\bf y}_r$ at the relay is given by
\begin{eqnarray}\label{y_r}
{{\bf y}}_r=\sum\limits_{i=1}^{K}{{\bf H}}_{i,r}{{\bf x}}_{i}+{{\bf n}}_r,
\end{eqnarray}
where ${{\bf H}}_{i,r}$ denotes the frequency-flat quasi-static $N \times M$ complex-valued channel matrix from source node $i$ to the relay and ${{\bf n}}_r$ denotes the $N\times 1$ additive white Gaussian noise (AWGN) with variance $\sigma_n^2$.

In the BC phase, upon receiving ${{\bf y}}_r$ in \eqref{y_r}, the relay processes it to obtain a mixed signal ${\bf x}_r$, and broadcasts to all the users. The transmitted signal ${\bf x}_r$ satisfies the power constraint of
\begin{equation}\label{Power_constraint_relay}
\textrm{tr}({\bf x}_r{\bf x}_r^H) \leq P_{\rm{r}},
\end{equation}
where $P_{\rm{r}}$ is the maximum transmission power allowed at the relay. Without loss of generality from the perspective of DoF analysis, we let $P_{\rm{s}}=P_{\rm{r}}=P$. The received signal at source node $i$ can be written as
\begin{eqnarray}\label{y_i}
{{\bf y}}_i={\bf G}_{r,i}{\bf x}_r+{{\bf n}}_i,
\end{eqnarray}
where ${{\bf G}}_{r,i}$ denotes the frequency-flat quasi-static $M \times N$ complex-valued channel matrix from relay to the source node $i$, and ${\bf n}_i$ denotes the AWGN at the source node $i$. Each user tries to obtain its desired signal from its received signal using its own transmit signal as side information.

It is assumed that the channel state information $\big\{{\bf H}_{i,r}, {\bf G}_{r,i}\big\}$ is perfectly known at all source nodes and the relay, following the convention in \cite{Lee,Wang3,Mu1,Wang4,Liu3,Liu5}. If the CSI is not perfectly known, the DoF will reduce. The discussion of the achievable DoF with imperfect CSI is beyond the scope of this paper. The entries of the channel matrices and those of the noise vectors $\big\{{\bf n}_r$, ${\bf n}_i\big\}$ are independent and identically distributed (i.i.d.) zero-mean complex Gaussian random variables with unit variance. Thus, each channel matrix is of full rank with probability $1$.

\section{Degrees of Freedom Upper Bound}
Let $R_{i,j}$ denote the information rate carried in $W_{i,j}$. Since we assume the noise is i.i.d. zero-mean complex Gaussian random variables with unit variance, the average received SNR of each link is $P$. We define the DoF of the transmission from source node $i$ to source node $j$ as
\begin{equation}\label{dof_source}
d_{i,j} \triangleq \lim\limits_{\textrm{SNR} \rightarrow \infty} \frac{R_{i,j}(\textrm{SNR})}{\textrm{log}(\textrm{SNR})}=\lim\limits_{P \rightarrow \infty} \frac{R_{i,j}(P)}{\textrm{log}(P)}.
\end{equation}

\textit{Theorem 1}: The DoF for the $K$-user MIMO Y channel is piece-wise upper-bounded by
\begin{align}\label{dof_KY_upper_information}
d_{total}^{u}=\left\{
                    \begin{array}{ll}
                      2N, & \hbox{$\frac{N}{M} \in \big(0, \frac{2K^2-2K}{K^2-K+2}\big]$;} \\
                      \frac{2\beta K(K-1)M}{K(K-1)+\beta (\beta-1)}, & \hbox{$\frac{N}{M} \in \big(\frac{\beta(K(K-1)+(\beta-1)(\beta-2))}{K(K-1)+\beta (\beta-1)}, \beta\big]$;} \\
                      \frac{2K(K-1)N}{K(K-1)+\beta (\beta-1)}, & \hbox{$\frac{N}{M} \in \big(\beta, \frac{(\beta+1)(K(K-1)+\beta(\beta-1))}{K(K-1)+(\beta+1)\beta}\big]$;} \\
                      KM, & \hbox{$\frac{N}{M} \in \big(\frac{K^2-3K+3}{K-1}, +\infty \big)$.}
                    \end{array}
                  \right.
\end{align}
where $\beta \in \{2,3,4,\cdots,K-2\}$.
\begin{proof}
\footnote{This proof is followed by the method of \cite{Wang3,Wang4}.}Consider that each source node can decode the $K-1$ intended messages with its own $K-1$ messages as side information. Then, if a genie provides that side information to the relay, the relay is able to decode the messages desired at that source node and the sum rate will not decrease.

We first consider the case when $\frac{N}{M} \in \big(0, \frac{2K^2-2K}{K^2-K+2}\big]$. As illustrated in Fig. \ref{Fig_genie_Y_1}, we provide the genie information ${\cal G}_1=\{W_{i,j} \mid i=1,2,\cdots,K-1; j=i+1,i+2,\cdots,K\}$ to the relay. We can obtain \eqref{genie_Y_1} at the bottom of the next page, where $n$ is the number of time slots, and $\epsilon(n)$ represents that $\lim\limits_{n \rightarrow \infty} \frac{\epsilon(n)}{n}=0$ and $X_i$ represents all the messages transmitted from source node $i$.

\begin{figure*}[hb]
\hrule
\begin{align}\nonumber
&n(R_{i+1,i}+R_{i+2,i}+\cdots+R_{K,i})\\\nonumber
\leq & I(W_{i+1,i},W_{i+2,i},\cdots,W_{K,i};Y_i^n \mid W_{i,1}, W_{i,2}, \cdots, W_{i,i-1}, W_{i,i+1},\cdots, W_{i,K})+\epsilon(n)\\\nonumber
\leq & I(W_{i+1,i},W_{i+2,i},\cdots,W_{K,i};Y_r^n \mid W_{i,1}, W_{i,2}, \cdots, W_{i,i-1}, W_{i,i+1},\cdots, W_{i,K})+\epsilon(n)\\\nonumber
\leq & I(W_{i+1,i},W_{i+2,i},\cdots,W_{K,i};Y_r^n, {\cal G}_1\mid W_{i,1}, W_{i,2}, \cdots, W_{i,i-1}, W_{i,i+1},\cdots, W_{i,K})+\epsilon(n)\\\nonumber
= & I(W_{i+1,i},W_{i+2,i},\cdots,W_{K,i};{\cal G}_1\mid W_{i,1}, W_{i,2}, \cdots, W_{i,i-1}, W_{i,i+1},\cdots, W_{i,K})\\\nonumber
&~+I(W_{i+1,i},W_{i+2,i},\cdots,W_{K,i};Y_r^n \mid {\cal G}_1, W_{i,1}, W_{i,2}, \cdots, W_{i,i-1}, W_{i,i+1},\cdots, W_{i,K})+\epsilon(n)\\\nonumber
= & I(W_{i+1,i},W_{i+2,i},\cdots,W_{K,i};Y_r^n \mid {\cal G}_1, W_{i,1}, W_{i,2}, \cdots, W_{i,i-1}, W_{i,i+1},\cdots, W_{i,K})+\epsilon(n)\\\label{genie_Y_1}
= & I(W_{i+1,i},W_{i+2,i},\cdots,W_{K,i};Y_r^n \mid {\cal G}_1, W_{i,1}, W_{i,2}, \cdots, W_{i,i-1})+\epsilon(n)
\end{align}
\vspace{-0.5cm}
\end{figure*}

Adding \eqref{genie_Y_1} from $i=1$ to $K-1$, we can obtain
\begin{align}\nonumber
&n(\sum\limits_{i=1}^{K-1} \sum\limits_{j=i+1}^{K} R_{j,i})\\\nonumber
\leq & I(\{W_{j,i} \mid i=1,2,\cdots,K-1; j=i+1,i+2,\cdots,K\};\\\nonumber
&~Y_r^n \mid {\cal G}_1)+\epsilon(n)\\\nonumber
\leq & h(Y_r^n \mid {\cal G}_1)+\epsilon(n)\\\label{genie_Y_1_sum}
\leq & nN\textrm{log}P+\epsilon(n)
\end{align}
Dividing $n\textrm{log}P$ to both side of \eqref{genie_Y_1_sum} and letting $n\rightarrow \infty$ and $P\rightarrow \infty$, we can obtain the total DoF upper bound as
\begin{align}
d_{total}=\sum\limits_{i=1}^{K} \sum\limits_{j=1}^K d_{i,j} \leq 2N.
\end{align}

\begin{figure}[t]
\begin{centering}
\includegraphics[scale=0.3]{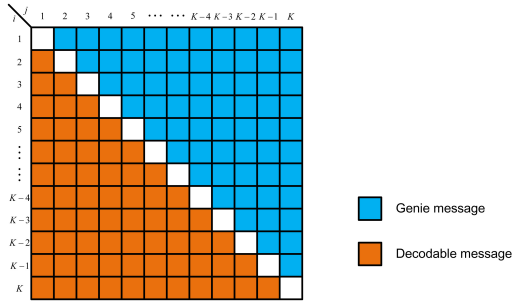}
\vspace{-0.1cm}
 \caption{Illustration for the genie information and the decodable messages at the relay for the $K$-user MIMO Y channel when $\frac{N}{M} \in \big(0, \frac{2K^2-2K}{K^2-K+2}\big]$}\label{Fig_genie_Y_1}
\end{centering}
\vspace{-0.3cm}
\end{figure}

Next, we consider the case when $\frac{N}{M} \in \big(\beta, \frac{(\beta+1)(K(K-1)+\beta(\beta-1))}{K(K-1)+(\beta+1)\beta}\big]$ for each $\beta \in \{2,3,4,\cdots,K-2\}$. As illustrated in Fig. \ref{Fig_genie_Y_beta}, we provide the genie information ${\cal G}_2=\{W_{i,j} \mid i=1,2,\cdots,K-\beta; j=i+1,i+2,\cdots,K\}$ to the relay. Similar to \eqref{genie_Y_1}, we can obtain
\begin{align}
&n(R_{i+1,i}+R_{i+2,i}+\cdots+R_{K,i})\\\label{genie_Y_beta}
\leq & I(W_{i+1,i},\cdots,W_{K,i};Y_r^n \mid {\cal G}_2, W_{i,1}, \cdots, W_{i,i-1})+\epsilon(n)
\end{align}
Adding \eqref{genie_Y_beta} from $i=1$ to $K-\beta$, we can obtain \eqref{genie_Y_beta_sum} at the bottom of the next page.
\begin{figure*}[hb]
\hrule
\begin{align}\nonumber
&n(\sum\limits_{i=1}^{K-\beta} \sum\limits_{j=i+1}^{K} R_{j,i})\\\nonumber
\leq & I(\{W_{j,i} \mid i=1,2,\cdots,K-\beta; j=i+1,i+2,\cdots,K\};Y_r^n \mid {\cal G}_2)+\epsilon(n)\\\nonumber
= & h(Y_r^n \mid {\cal G}_2)-h(Y_r^n \mid {\cal G}_2,\{W_{j,i} \mid i=1,2,\cdots,K-\beta; j=i+1,i+2,\cdots,K\})+\epsilon(n)\\\nonumber
= & h(Y_r^n \mid {\cal G}_2)-h(Y_r^n \mid X_1^n,X_2^n,\cdots,X_{K-\beta}^n,\{W_{j,i} \mid i\in[1,K-\beta]; i\in \textbf{Z}; j\in [K-\beta+1,K];j\in \textbf{Z}\})+\epsilon(n)\\\nonumber
= & h(Y_r^n \mid {\cal G}_2)-h(X_{K-\beta+1}^n,X_{K-\beta+2}^n,\cdots,X_{K}^n \mid \{W_{j,i} \mid i\in[1,K-\beta]; i\in \textbf{Z}; j\in [K-\beta+1,K];j\in \textbf{Z}\})\\\nonumber
&~+n\epsilon(\textrm{log}P)+\epsilon(n)\\\nonumber
= & h(Y_r^n \mid {\cal G}_2)-H(\{W_{i,j} \mid i \in [K-\beta+1,K];i\in \textbf{Z}; j \in [K-\beta+1,K];j\in \textbf{Z};j \neq i\})+n\epsilon(\textrm{log}P)+\epsilon(n)\\\label{genie_Y_beta_sum}
\leq & nN\textrm{log}P-n(\{R_{i,j} \mid i \in [K-\beta+1,K];i\in \textbf{Z}; j \in [K-\beta+1,K];j\in \textbf{Z};j \neq i\})+n\epsilon(\textrm{log}P)+\epsilon(n)
\end{align}
\vspace{-0.5cm}
\end{figure*}
Then we can obtain
\begin{align}\nonumber
&n(\sum\limits_{i=1}^{K-1} \sum\limits_{j=i+1}^{K} R_{j,i}+\sum\limits_{i=K-\beta+1}^{K-1} \sum\limits_{j=i+1}^{K} R_{i,j})\\\label{R}
\leq & nN\textrm{log}P+n\epsilon(\textrm{log}P)+\epsilon(n).
\end{align}

We can obtain similar equations to \eqref{R} by replacing any $\beta$ source nodes to $K-\beta+1,K-\beta+2,\cdots,K$. Then dividing $n\textrm{log}P$ to both side of \eqref{R} and letting $n\rightarrow \infty$ and $P\rightarrow \infty$, we can obtain the total DoF upper bound as
\begin{align}
d_{total}=\sum\limits_{i=1}^{K} \sum\limits_{j=1}^K d_{i,j} \leq \frac{2K(K-1)N}{K(K-1)+\beta(\beta-1)}.
\end{align}

\begin{figure}[t]
\begin{centering}
\includegraphics[scale=0.3]{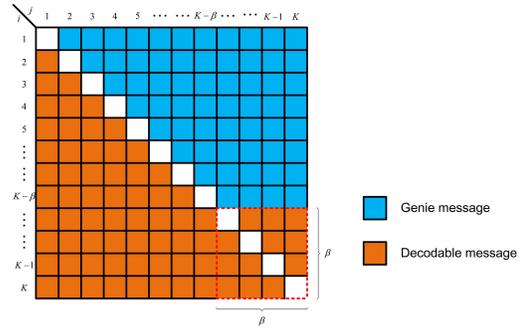}
\vspace{-0.1cm}
 \caption{Illustration for the genie information and the decodable messages at the relay for the $K$-user MIMO Y channel when $\frac{N}{M} \in \big(\beta, \frac{(\beta+1)(K(K-1)+(\beta)(\beta-1))}{K(K-1)+(\beta+1)\beta}\big]$}\label{Fig_genie_Y_beta}
\end{centering}
\vspace{-0.3cm}
\end{figure}

Third, we consider the case when $\frac{N}{M} \in \big(\frac{\beta(K(K-1)+(\beta-1)(\beta-2))}{K(K-1)+\beta (\beta-1)}, \beta\big]$. We prove this by contradiction. If $\frac{N}{M}  \in \big(\frac{\beta(K(K-1)+(\beta-1)(\beta-2))}{K(K-1)+\beta (\beta-1)}, \beta\big]$ and $d_{total}^a > \frac{2\beta K(K-1)M}{K(K-1)+\beta (\beta-1)}$, where $d_{total}^a$ represents the achivable total DoF, then we increase $N$ to $N_1$ such that $\frac{N_1}{M} =\beta$. Utilizing the antenna deactivation, $d_{total}^a > \frac{2\beta K(K-1)M}{K(K-1)+\beta (\beta-1)}=\frac{2 K(K-1)N_1}{K(K-1)+\beta (\beta-1)}$ can be achieved. However, when $\frac{N_1}{M} =\beta$, the DoF upper bound is $\frac{2 K(K-1)N_1}{K(K-1)+\beta (\beta-1)}$. There is a contradiction. Hence, the DoF upper bound of the case when $\frac{N}{M} \in \big(\frac{\beta(K(K-1)+(\beta-1)(\beta-2))}{K(K-1)+\beta (\beta-1)}, \beta\big]$ is $\frac{2\beta K(K-1)M}{K(K-1)+\beta (\beta-1)}$.

Finally, we consider the case when $\frac{N}{M} \in \big(\frac{K^2-3K+3}{K-1}, +\infty \big)$. In this case, we notice that the DoF per user could not be larger than $M$. Thus, $KM$ is the DoF upper bound for the case when $\frac{N}{M} \in \big(\frac{K^2-3K+3}{K-1}, +\infty \big)$.
\end{proof}

\section{Generalized Signal Alignment}
\subsection{Basic principles}
We rewrite the received signal \eqref{y_r} at the relay during the MAC phase as
\begin{align}\label{y_r_align}
{{\bf y}}_r=\sum\limits_{i=1}^K {\bf H}_{i,r}{\bf V}_i\textbf{s}_i+{\bf n}_r.
\end{align}
Define $\textbf{s}_{\oplus}$ as a network-coded symbol vector obtained by stacking the $\{{\bf s}_{i,j}+{\bf s}_{j,i},~\forall j > i\}$ by row. When $N \geq 2M$, the network-coded symbol vector $\textbf{s}_{\oplus}$ cannot be obtained directly by designing the precoding matrices ${\bf V}_{i,j}$ and ${\bf V}_{j,i}$ \cite{Lee1}. Instead, the joint design of the source precoding matrices and relay compression matrix should be considered \cite{Liu5}. We introduce a $\frac{d_{total}}{2} \times N$ full-rank compression matrix $\bf P$ to compress the signal ${\bf y}_r$ as
\begin{align}\label{y_r_projection}
\hat{{\bf y}}_r={{\bf P}}{{\bf y}}_r=\sum\limits_{i=1}^K {\bf P}{\bf H}_{i,r}{\bf V}_i\textbf{s}_i+{\bf P}{\bf n}_r,
\end{align}
so that the signals can be aligned as
\begin{align}\label{GSA_constraints}
{\bf P}{\bf H}_{i,r}{\bf V}_{i,j}={\bf P}{\bf H}_{j,r}{\bf V}_{j,i} \triangleq {\bf B}_{i,j}.
\end{align}
Here, ${\bf V}_{i,j}$ should all be of full rank. If not, the network-coded symbol vector $\textbf{s}_{\oplus}$ cannot be decoded and the DoF will reduce. The compressed signal $\hat{{\bf y}}_r$ can be rewritten as
\begin{align}\label{y_r_projection}
\hat{{\bf y}}_r={\bf B}{\bf s}_{\oplus}+{\bf P}{\bf n}_r,
\end{align}
where $\bf B$ is a matrix obtained by stacking the ${\bf B}_{i,j}$ by column.

We refer to the above condition \eqref{GSA_constraints} as \textit{generalized signal alignment equation}.

\textit{Remark 1}: Since the entries of all channel matrices are independently Gaussian, the probability that a basis vector in the intersection space of one pair of source nodes¡¯ channel matrices lies in the intersection space of another pair is zero \cite{Lee1}. Thus, $\bf B$ is of full-rank with probability 1, which guarantees the decodability of $\textbf{s}_{\oplus}$ at the relay.

\textit{Theorem 2}: The GSA equation \eqref{GSA_constraints} holds if and only if the following two conditions hold:.
\begin{enumerate}
  \item There are at least $\frac{d_{total}}{2}-2M+d_{i,j}$ row vectors of $\bf P$ lie in the left null space of $\left[{\bf H}_{i,r}~-{\bf H}_{j,r}\right]$ for any source pair $(i,j)$.
  \item For any source pair $(i,j)$,
\begin{align}\label{GSA_v_a}
\left[\begin{array}{c}
        {\bf V}_{i,j} \\
        {\bf V}_{j,i}
      \end{array}
\right] \subseteq \textbf{Null}~\left[{\bf P}{\bf H}_{i,r}~-{\bf P}{\bf H}_{j,r}\right].
\end{align}
\end{enumerate}

\begin{proof}
First, we prove its necessity. For any source pair $(i,j)$, the above alignment condition in \eqref{GSA_constraints} can be rewritten as
\begin{align}\label{GSA_2}
\left[{\bf P}{\bf H}_{i,r}~-{\bf P}{\bf H}_{j,r}\right]
\left[ \begin{array}{c}
        {\bf V}_{i,j} \\
        {\bf V}_{j,i}
      \end{array}
\right]&=0
\end{align}
or
\begin{align}\label{GSA_3}
{\bf P}\left[{\bf H}_{i,r}~-{\bf H}_{j,r}\right]
\left[ \begin{array}{c}
        {\bf V}_{i,j} \\
        {\bf V}_{j,i}
      \end{array}
\right]&=0.
\end{align}

Define ${\bf A}_{i,j}=\left[{\bf P}{\bf H}_{i,r}~-{\bf P}{\bf H}_{j,r}\right]$. Clearly, the dimension of ${\bf A}_{i,j}$ is $\frac{d_{total}}{2} \times 2M$. For \eqref{GSA_2} to hold, one must have that the dimension of the null space of the matrix ${\bf A}_{i,j}$ should be greater than $d_{i,j}$ for signal alignment \cite{Lee}. That is
\begin{align}\nonumber
d_{i,j} &\leq 2M-\textrm{rank}({\bf A}_{i,j})
\end{align}
or equivalently
\begin{align}\label{rank_aij}
\textrm{rank}({\bf A}_{i,j}) &\leq 2M-d_{i,j}.
\end{align}
Hence, from \eqref{GSA_3}, at least $\frac{d_{total}}{2}-2M+d_{i,j}$ row vectors of $\bf P$ should lie in the left null space of $\left[{\bf H}_{i,r}~-{\bf H}_{j,r}\right]$.

Once we can find a compression matrix $\bf P$ to satisfy the above condition, we design the precoding matrices ${\bf V}_{i,j}$ and ${\bf V}_{j,i}$ for all source pairs $(i,j)$ as in 2) to meet the GSA equation \eqref{GSA_constraints}.

Then we prove its sufficiency. For the GSA equation \eqref{GSA_constraints} holds, the precoding matrices must satisfy \eqref{GSA_v_a}. Once \eqref{GSA_v_a} is satisfied, we can obtain that the dimension of the null space of $\left[{\bf P}{\bf H}_{i,r}~-{\bf P}{\bf H}_{j,r}\right]$ must be no less than $d_{i,j}$, which indicates that $2M-\textrm{rank}(\left[{\bf P}{\bf H}_{i,r}~-{\bf P}{\bf H}_{j,r}\right])$ must be no less than $d_{i,j}$ i.e. \eqref{rank_aij}. Since $\left[{\bf H}_{i,r}~-{\bf H}_{j,r}\right]$ is an $N \times 2M$ matrix of full rank with probability 1, then the only way to reduce its rank is that there are at least $\frac{d_{total}}{2}-2M+d_{i,j}$ row vectors of $\bf P$ lie in the left null space of $\left[{\bf H}_{i,r}~-{\bf H}_{j,r}\right]$.
\end{proof}

In our proposed GSA, the signals to be exchanged do not align directly in the subspace observed by the relay. Instead, they are aligned in a compressed subspace after the compression of $\bf P$ at the relay.

\subsection{Comparison to the existing transmission framework}
Previously, there are two main transmission frameworks to analyze the DoF when $N \geq 2M$. The first method is proposed in \cite{Mu1}, named as \textit{signal group based alignment}. The second is proposed in \cite{Wang3}, named as \textit{signal pattern}. The main idea of the two methods is that the precoding matrices at each source node are first constructed under certain rules, such as group or pattern. The signal received at the relay is then proved to be decodable into $\textbf{s}_{\oplus}$ with specific signal processing. In our proposed GSA, we first design the compression matrix $\bf P$ at the relay, and then construct the precoding matrices at each source node. It can be seen that the goal of both the previous methods and our method is to design the precoding matrices $\{{\bf V}_{i,j}\}$ and the compression matrix $\bf P$ to form the network-coded symbol vector $\textbf{s}_{\oplus}$. It is worth mentioning that these three methods all satisfy the GSA equation \eqref{GSA_constraints}.

The main difference between the previous transmission frameworks and our GSA transmission framework is that we reverse the design order of $\bf P$ at the relay node and $\{{\bf V}_{i,j}\}$ for each source node. Note that $\bf P$ is a common compression matrix for processing the signal at the relay and $\{{\bf V}_{i,j}\}$ is a set of private matrices designed for each source node. The existing transmission framework \cite{Mu1} and \cite{Wang3} first design the private parameters at each source node. This will lock the pattern of the signals received at the relay. On the other hand, in our proposed GSA, the common parameter is first designed and then the private parameters. As a result, the signal received at the relay does not need to have any pattern, which is a more general transmission framework for alignment. This leads the increase of the DoF compared to the existing work.

\section{Achievable DoF of the $K$-user MIMO Y channel}
Define a coordinate $Q = (y,z)$ where $y$ represents the antenna configuration $\frac{N}{M}$ and $z$ represents DoF value at $\frac{N}{M}$.

\textit{Lemma 1}: If the point $Q=(\alpha, d_0 M)$ is achievable for some $\alpha >0$ and $d_0 > 0$, then the DoF $d_0 M$ is achievable when $\frac{N}{M}>\alpha$ and $\frac{d_0 N}{\alpha}$ is achievable when $\frac{N}{M}<\alpha$.
\begin{proof}
When $\frac{N}{M}>\alpha$, let the relay node only utilize $\alpha M$ antennas. Then this turns to the case of $Q$, yielding the DoF of $d_0 M$. When $\frac{N}{M}<\alpha$, let each source node only utilize $\frac{N}{\alpha}$ antennas. Then this turns to the case of $Q$, yielding the DoF of $\frac{d_0 N}{\alpha}$.
\end{proof}

\textit{Theorem 3}:  \footnote{In \cite{Liu5}, only $\beta = K-2$ is considered.}For the $K$-user MIMO Y channel, the achievable points are given by \eqref{P} at the bottom of the next page, where $\beta \in \{2,3,4,\cdots,K-2\}$.
\begin{figure*}[hb]
\hrule
\begin{subequations}\label{P}
\begin{align}
&Q_1=\left(\frac{2K^2-2K}{K^2-K+2},\frac{(4K^2-4K)M}{K^2-K+2}\right)\\
&Q_{\beta}=\left(\beta+\frac{2K(K-1)}{\big(2+K(K-1)-\beta(\beta-1)\big)\binom{K}{\beta}},\frac{4K(K-1)M}{2+K(K-1)-\beta(\beta-1)}\right)
\end{align}
\end{subequations}
\vspace{-0.5cm}
\end{figure*}
\begin{proof}
$Q_1$ is proved to be achievable in \cite{Lee}. In what follows, we prove the DoF achievability of $Q_{\beta}$ when $\beta \geq 2$. Note that the abscissa of $Q_{\beta}$ is located in the interval $[\beta,\beta+1]$.

For the symmetry of each source node, we assume that $d_{i,j}$ for any pair $(i,j)$ is the same and given by $x$, then $d_{total}$ is $K(K-1)x$. For $\frac{N}{M} \geq \beta$, then the dimension of the left null space of any $\beta$-combining channel matrices $\left[{\bf H}_{\gamma_{1},r}~{\bf H}_{\gamma_{2},r},\cdots,{\bf H}_{\gamma_{\beta},r} \right]$, consisting of arbitrary $\beta$ channel matrices with $\left\{\gamma_{1},\gamma_{2},\cdots,\gamma_{\beta}\right\} \subseteq \left\{1,2,3,\cdots,K\right\}$, must be greater than zero. We design $q$ row vectors of the matrix $\bf P$ such that they are located in the left null space of the matrix $\left[{\bf H}_{\gamma_{1},r}~{\bf H}_{\gamma_{2},r},\cdots,{\bf H}_{\gamma_{\beta},r} \right]$. Clearly, we can obtain that $q \leq N-\beta M$. Then we will show the number of row vectors which are located in the left null space of $\left[{\bf H}_{i,r}~-{\bf H}_{j,r}\right]$. Consider the following $\beta$-combining channel matrix.
\begin{align}\label{P_i_j_number}
\left[{\bf H}_{i,r}~{\bf H}_{j,r}~~~~\underbrace{\square~~~~\square~~\cdots~~\square}_{\beta-2}\right]
\end{align}
We can find that there are total $\binom{K-2}{\beta -2}$ different cases. Hence, there are $\binom{K-2}{\beta -2}q$ row vectors which are located in the left null space of $\left[{\bf H}_{i,r}~-{\bf H}_{j,r}\right]$\footnote{The dimension of the left null space of $\left[{\bf H}_{i,r}~-{\bf H}_{j,r}\right]$ is equal to that of $\left[{\bf H}_{i,r}~{\bf H}_{j,r}\right]$}. Define the number of the row vectors located in the left null space of $\left[{\bf H}_{i,r}~-{\bf H}_{j,r}\right]$ as $p_{i,j}$, we can obtain
\begin{align}\label{N_M_p}
N-\beta M \geq q \geq \frac{p_{i,j}}{\binom{K-2}{\beta -2}}.
\end{align}

On the other hand, the number of the rows of $\bf P$ is $\frac{d_{total}}{2}$ and the total cases of $\beta$-combining channel matrices is $\binom{K}{\beta}$, which can be written as
\begin{align}\label{p_q_dtotal_1}
\frac{d_{total}}{2} \geq \binom{K}{\beta}q \geq \frac{\binom{K}{\beta}}{\binom{K-2}{\beta -2}}p_{i,j},
\end{align}
which is equivalent to
\begin{align}\label{p_q_dtotal_2}
p_{i,j} \leq \frac{d_{total}\binom{K-2}{\beta -2}}{2\binom{K}{\beta}}.
\end{align}
From \textit{Throrem 2}, we have
\begin{align}\label{p_constraints_1}
p_{i,j} \geq \frac{d_{total}}{2}-2M+d_{i,j}.
\end{align}
Combining \eqref{p_q_dtotal_2} and \eqref{p_constraints_1}, the relationship between $d_{i,j}$ and $M$ can be derived as
\begin{align}\nonumber
x&=d_{i,j} \leq 2M-\frac{d_{total}}{2}+p_{i,j}\\\nonumber
& \leq 2M-\frac{d_{total}}{2}+\frac{d_{total}\binom{K-2}{\beta -2}}{2\binom{K}{\beta}}\\
&\leq 2M-\frac{K(K-1)x}{2}+\frac{K(K-1)x\binom{K-2}{\beta -2}}{2\binom{K}{\beta}}\\\nonumber
x &\leq \frac{4\binom{K}{\beta}M}{2\binom{K}{\beta}+K(K-1)\binom{K}{\beta}-K(K-1)\binom{K-2}{\beta-2}}\\\nonumber
&\leq \frac{4\binom{K}{\beta}M}{2\binom{K}{\beta}+K(K-1)\binom{K}{\beta}-K(K-1)\big[\binom{K}{\beta}\frac{\beta (\beta-1)}{K(K-1)}\big]}\\
&=\frac{4M}{2+K(K-1)-\beta (\beta-1)}
\end{align}

Take $x=\frac{4M}{2+K(K-1)-\beta (\beta-1)}$, with \eqref{p_constraints_1} and \eqref{N_M_p}, we can obtain
\begin{align}\nonumber
N &\geq \beta M+\frac{p_{i,j}}{\binom{K-2}{\beta -2}}\\\nonumber
&\geq \beta M+\frac{\frac{d_{total}}{2}-2M+d_{i,j}}{\binom{K-2}{\beta -2}}\\\nonumber
&= \beta M+\frac{\frac{K(K-1)x}{2}-2M+x}{\binom{K-2}{\beta -2}}\\
&= \beta M+\frac{2K(K-1)M}{\big(2+K(K-1)-\beta(\beta-1)\big)\binom{K}{\beta}}
\end{align}

Hence, the total DoF of $K(K-1)x=\frac{4K(K-1)M}{2+K(K-1)-\beta (\beta-1)}$ can be achieved when $N \geq \beta M+\frac{2K(K-1)M}{\big(2+K(K-1)-\beta(\beta-1)\big)\binom{K}{\beta}}$. Taking the equality, we have the corner point $Q_{\beta}$.

Note that if the number of antennas after deactivation is not an integer but a fraction $\frac{s}{t}$, then we can use the method of $t$-symbol extensions to achieve the total DoF \cite{Liu5}. The theorem is thus proved.
\end{proof}

\begin{figure}[t]
\begin{centering}
\includegraphics[scale=0.4]{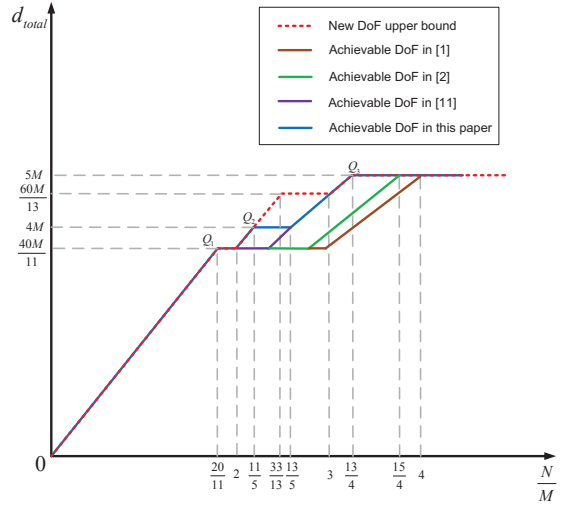}
\vspace{-0.1cm}
 \caption{New DoF upper bound and its achievability when $K=5$}\label{DoF_Y_5}
\end{centering}
\vspace{-0.3cm}
\end{figure}

\textit{Remark 2}: With \textit{Lemma 1} and \textit{Theorem 3}, we can easily obtain the achievable DoF at all possible $\frac{N}{M}$ with GSA. In specific, we can obtain $Q_1=\left(\frac{2K^2-2K}{K^2-K+2},\frac{(4K^2-4K)M}{K^2-K+2}\right)$, $Q_{2}=(2+\frac{4}{K(K-1)},4M)$ and $Q_{K-2}=(\frac{K^2-3K+3}{K-1},KM)$. When $\frac{N}{M} \in \big(0, \frac{2K^2-2K}{K^2-K+2}\big]$, the achievable DoF is $2N$. When $\frac{N}{M} \in \big(\frac{2K^2-2K}{K^2-K+2}, 2\big]$, the achievable DoF is $\frac{(4K^2-4K)M}{K^2-K+2}$. When $\frac{N}{M} \in \big(2, 2+\frac{4}{K(K-1)}\big]$, the achievable DoF is $\frac{(2K^2-2K)N}{K^2-K+2}$. When $\frac{N}{M} \in \big(K-2, \frac{K^2-3K+3}{K-1}\big]$, the achievable DoF is $\frac{K(K-1)N}{K^2-3K+3}$. When $\frac{N}{M} \in \big(\frac{K^2-3K+3}{K-1}, +\infty\big]$, the achievable DoF is $KM$. Furthermore, from \textit{Theorem 1}, we find that the DoF upper bound under the antenna configuration $\frac{N}{M} \in \big(0, 2+\frac{4}{K(K-1)}\big] \cup \big[K-2, +\infty\big)$ can be achieved by GSA when $K>4$. Fig. \ref{DoF_Y_5} illustrates the new DoF upper bound and its achievability when $K=5$.

\textit{Remark 3}: If $K=4$, then only consider $\beta =2$ and the DoF upper bound can be achieved by GSA under the antenna configuration $\frac{N}{M} \in \big(0, +\infty\big)$, which meets the same conclusion of the $4$-user MIMO Y channel in \cite{Wang4}.

\section{Conclusion}
In this paper, we have analyzed the DoF for the $K$-user MIMO Y channel. A new DoF upper bound is derived. We have shown that under the antenna configuration $\frac{N}{M} \in \big(0, 2+\frac{4}{K(K-1)}\big] \cup \big[K-2, +\infty\big)$, the new DoF upper bound can be achieved by the proposed GSA. Despite the existing advances in studying the fundamental capacity limit of the $K$-user MIMO Y channel, this work provides a major step forward. The proposed generalized signal alignment is also a new approach of integrating interference alignment with physical layer network coding.

\bibliographystyle{IEEEtran}
\bibliography{IEEEabrv,reference}

\end{document}